%% file: main.tex
\documentclass[sigconf, screen]{acmart}
\usepackage{xcolor}

\input{meta/copyright}

\input{meta/acronyms}

\input{meta/packages}

\begin{document}

\input{meta/authors}
\input{meta/macro}
\input{sections/00_abstract}
\input{meta/metadata}

\maketitle

\input{sections/01_introduction}

\input{sections/02_related_work}

\input{sections/03_methodology}
\input{sections/04_findings}

\input{sections/06_discussion}
\input{sections/07_conclusion}

\input{sections/08_ack}

\bibliographystyle{ACM-Reference-Format}
\bibliography{ref}
\end{document}

%% file: meta/copyright.tex
\AtBeginDocument{%
  }

\copyrightyear{2025}
\acmYear{2025}
\setcopyright{rightsretained}
\acmConference[CHI EA '25]{Extended Abstracts of the CHI Conference on Human Factors in Computing Systems}{April 26-May 1, 2025}{Yokohama, Japan}
\acmBooktitle{Extended Abstracts of the CHI Conference on Human Factors in Computing Systems (CHI EA '25), April 26-May 1, 2025, Yokohama, Japan}\acmDOI{10.1145/3706599.3719993}
\acmISBN{979-8-4007-1395-8/2025/04}

%% file: meta/acronyms.tex
\usepackage[nolist]{acronym}
\begin{acronym}
    \acro{HCI}{Human-Computer Interaction}
    \acro{CSCW}{Computer-Supported Collaborative Work}
    \acro{CS}{Computer Science}
    \acro{LLM}{Large Language Model}
\end{acronym}

%% file: meta/packages.tex
\usepackage{amssymb}
\usepackage{makecell}
\usepackage{dirtytalk}
\usepackage{indentfirst}
\usepackage{url}

%% file: meta/authors.tex
\author{Xiaotian Su}
\email{xiaotian.su@inf.ethz.ch}
\orcid{0009-0004-0548-1576}
\affiliation{%
  \institution{ETH Zurich}
  \city{Zurich}
  \country{Switzerland}
}

\author{April Yi Wang}
\email{april.wang@inf.ethz.ch}
\orcid{0000-0001-8724-4662}
\affiliation{%
  \institution{ETH Zurich}
  \city{Zurich}
  \country{Switzerland}
}

\renewcommand{\shortauthors}{Su et al.}

%% file: meta/macro.tex
\newcommand{\AW}[1]{\textcolor{blue}{\textbf{*April*}: #1}}
\newcommand{\SW}[1]{\textcolor{purple}{\textbf{*Su*}: #1}}
\newcommand{\isay}[1]{\say{\textit{#1}}}

%% file: sections/00_abstract.tex
\begin{abstract}
Live coding is a pedagogical technique in which an instructor writes and executes code in front of students to impart skills like incremental development and debugging. 
Although live coding offers many benefits, instructors face many challenges in the classroom, like cognitive challenges and psychological stress, most of which have yet to be formally studied.
To understand the obstacles faced by instructors in CS classes, we conducted (1) a formative interview with five teaching assistants in exercise sessions and (2) a contextual inquiry study with four lecturers for large-scale classes.
We found that the improvisational and unpredictable nature of live coding makes it difficult for instructors to manage their time and keep students engaged, resulting in more mental stress than presenting static slides. We discussed opportunities for augmenting existing IDEs and presentation setups to help enhance live coding experience.
\end{abstract}

%% file: meta/metadata.tex
\newcommand{\sys}{SYSNAME}

\title{Live Coding in Programming Classes: Instructors' Perspectives}

\ccsdesc[500]{Human-centered computing~Empirical Studies in HCI
}

\keywords{live coding, programming education at scale}

%% file: sections/01_introduction.tex
\section{Introduction}
Live coding is a teaching method where instructors write code in real time in front of students while verbalizing their thoughts \cite{ace20-live-static, technique02}. This approach allows instructors to demonstrate debugging and problem-solving strategies as they occur \cite{cer18-role, cse13-livecoding}. Unlike static code examples, live coding presents code in a step-by-step manner so students know which line of code to focus on thus reducing their extraneous cognitive load \cite{ace20-live-static}. It also creates a more engaging, hands-on learning experience \cite{cse13-livecoding} and improves testing skills \cite{bennedsen2005revealing}, as reported by both students and instructors.

While live coding offers numerous benefits to students, our understanding of instructors' experiences with this technique remains limited \cite{iticse-21-literature-review}. For example, studies have suggested that instructors may experience ``stage fright'' -- anxiety about both the content and the style of their delivery \cite{07stage-fright}. By typing, debugging, and thinking aloud, instructors may experience more mental workload than in normal lectures \cite{iticse-21-literature-review}. However, to our knowledge, no study has explored the actual experiences and perceptions of instructors performing live coding. 
Specifically, there is little concrete evidence for how instructors perceive the benefits of live coding and the cognitive load it imposes on them, a gap that is highlighted by \citet{iticse-21-literature-review}. 
Addressing these questions is not only crucial for the overall impact of live coding in educational contexts but also for informing the design and development of innovative tools that can support instructors by reducing cognitive load, enhancing real-time interactions, and improving the overall efficacy of live coding sessions.

To explore the challenges associated with live coding, we propose the following research questions: (1) What motivates instructors to use live coding in their teaching? (2) How do instructors approach and execute live coding in the classroom? (3) What obstacles or barriers do instructors encounter when teaching programming through live coding? Through two complementary studies--formative interviews with five teaching assistants (TAs) and contextual inquiries with four lecturers—-we aim to provide a holistic understanding of live coding by exploring both reflective and contextual perspectives, thereby deepening insights into instructors' perceptions of this teaching practice.
We found that although instructors agree that live coding can provide adaptive content to engage students, it is a time consuming activity in terms of both preparation and presentation. The improv nature further adds burden to the instructors' mental load.
Based on these insights, we present design implications on the development of tools that provide real-time scaffolding to improve the effectiveness of live coding in educational settings.

%% file: sections/02_related_work.tex
\section{Related Work}
\subsection{Live Coding for Students in CS Classrooms}
Although live coding has not been conclusively shown to significantly improve measurable student learning outcomes \cite{ace20-live-static, icer23-empirical, cse13-livecoding}, this format offers many benefits to the teaching and learning experience compared to teaching programming with static code examples.
Firstly, live coding can reduce the extraneous cognitive load on students' working memory by presenting code in a step-by-step manner to guide students' focus \cite{ace20-live-static}. 
Secondly, live coding allows students to be actively engaged during the presentation, students appreciate they can be more involved in writing code along with the instructor in live coding sessions \cite{ace21-qualitative}
Thirdly, watching instructors make mistakes shows learners that it is alright to make mistakes \cite{brown2018ten}
Finally, when instructors think aloud during live coding, they reveal their thought processes and the reasoning behind their decisions. This approach not only makes abstract concepts more accessible but also provides students with insight into problem-solving strategies \cite{cscw21-towards, ace21-qualitative, bennedsen2005revealing}. Discussing code line by line further helps clarify complex programming concepts, which students often find challenging \cite{cscw21-towards, ace21-qualitative, bennedsen2005revealing}.

Despite the benefits, live coding presents several challenges that can hinder students' learning experiences and engagement. While it is preferred for explaining complex concepts, students perceive it as inefficient for simpler topics, where static code examples are more effective and time-saving \cite{ace21-qualitative}. 
Additionally, the dynamic nature of live coding complicates note-taking, as the program evolves continuously during the session \cite{ace20-live-static, icer23-empirical}.Students have expressed difficulty in mirroring the instructor’s edits in their notes, as it is easier for the instructor to modify code in real time.
Furthermore, instructors typically share only the final solutions, omitting the intermediate steps and thought processes that led to the result. This lack of captured progression reduces the value of the demonstration, making it harder for students to recall the solution's development after class \cite{sigcse19-coding}. Moreover, students who miss a session cannot benefit from the live coding experience, as it is a one-time performance.

\subsection{Challenges for Performing Live Coding}
\label{sec:rw-challenges}
While the benefits of live coding for students have been increasingly recognized, our understanding of instructors' experiences with this technique remains limited. Existing research predominantly focuses on student learning outcomes, leaving a significant gap in our knowledge about how instructors perceive and manage the challenges of live coding in practice \cite{iticse-21-literature-review}. This section explores the potential obstacles faced by instructors, drawing upon literature from related fields to hypothesize about the difficulties encountered during live coding sessions.

One major challenge likely faced by instructors is the high mental workload. Mental workload, defined as the cognitive resources a task demands from an individual \cite{MIYAKE2001233}, can be significantly elevated during live coding for two main reasons. First, the cognitive burden is increased due to the think-aloud component.
The think-aloud method is commonly employed by usability research to capture participants' thought processes \cite{psy-1980verbal}, it was demonstrated that this verbalization increases cognitive load compared to silent task performance \cite{09TF-scrutinising}.
Instructors verbalize their reasoning and actions during live coding, this kind of verbalization corresponds to the most complex level (level three) of verbalization according to \citet{psy-1980verbal}. This cognitive load is further exacerbated when instructors simultaneously address student questions \cite{09TF-scrutinising}.
Secondly, the inherent multitasking nature of live coding—simultaneously writing code, verbalizing thoughts, and interacting with students—can dramatically increase mental strain.
Multitasking has been shown to reduce efficiency significantly due to task-switching costs \cite{elsevier-03task-switch}, and high mental workloads during multitasking can lead to task management errors \cite{routledge-21engineering-psychology}. For example, \cite{vrst22-assessment} found that instructors reported greater workloads and lower communication quality when managing multiple student interactions compared to one-on-one settings.

In addition to cognitive challenges, instructors in classrooms may experience psychological stress. Stage fright, a form of social anxiety, stems from fears of making mistakes, forgetting material, or losing focus, leading to physical and emotional symptoms \cite{07stage-fright, marshall1994social}. This anxiety is heightened by the pressure to convey complex material in real time \citep{mead1934mind}, making live coding especially stressful. Despite these challenges derived from literature, the actual experiences and perceptions of instructors regarding live coding remain largely unexplored. This paper addresses this gap by examining instructors' motivations to use live coding, strategies they employ to engage students, and obstacles they encounter, aiming to inform the development of tools to support effective live coding.

%% file: sections/03_methodology.tex
\section{Methodology}
\begin{table*}[h!]
\centering
\begin{tabular}{cccccccccc}
\toprule
\textbf{PID} & \textbf{Age Range} & \textbf{Gender} & \textbf{Profession} & \textbf{\makecell{Teaching \\(\# years)}} & \textbf{\makecell{Class size \\(\# students)}} & \multicolumn{3}{c}{\textbf{\makecell{Students}}} \\ \cmidrule{7-9}
 & & & & & & H & U & C \\ \midrule
P1 & 20-25 & F & BSc Student & One & 20-30 & \checkmark & \checkmark & \\
P2 & 20-25 & M & \makecell{MSc Student, part-time lecturer} & Three & 20-30 & \checkmark & \checkmark & \\
P3 & 26-30 & M & PhD Student & Five & 20-30 & & \checkmark & \\
P4 & 20-25 & M & PhD Student & Five & 30-50 & & \checkmark & \checkmark \\
P5 & 26-30 & M & PhD Student & Six & 20-30 & \checkmark & \checkmark & \\
P6 & 30-35 & M & Full-time lecturer & Five & 300-400 & & \checkmark & \\
P7 & 40-45 & M & Full-time lecturer & Five & 50-150 & & \checkmark & \checkmark \\
P8 & 45-50 & M & Full-time lecturer & Eight & 100-150 & & \checkmark & \checkmark \\
\bottomrule
\end{tabular}
\caption{Demographics data of participants. H: High school students, U: University students, C: Continuing education.}
\label{tab:dmg-ta}
\end{table*}

We initially simulated live coding in a lab where participants solved a coding problem while explaining to the researchers. 
However, this setup failed to replicate key classroom aspects. 
Participants missed the stress of a real audience, lacked time constraints, and did not face the pressure of managing multiple concepts. 
The task’s low cognitive load, such as copy-pasting code from prepared notes, did not reflect the complexity of real live coding.
The environment also lacked interactivity, with no student input or unexpected challenges to address, resulting in minimal debugging. 

To overcome these limitations and ensure a realistic understanding of live coding practices, we conducted two complementary studies. First, we conducted formative interviews with TAs (N=5), focusing on their experiences and perceptions of live coding in small exercise sessions. Second, we carried out contextual inquiries with lecturers (N=4) by observing their live coding practices in large-scale lectures. 
Table \ref{tab:dmg-ta} reported the demographics data of participants from the two studies.
We captured a holistic view of live coding by combining both reflective and contextual perspectives, enabling us to identify common themes, challenges, and unique insights across different instructional roles and settings.

\subsection{Formative Study}
Five TAs (4M, 1F) with computer science backgrounds and teaching experience were recruited through campus posters.
They had experience teaching 
programming exercise sessions in high school or university settings, with class sizes ranging from 15 to 30 students. 
Programming languages taught included XLogo, Python, Java, and C++.  
The interview sessions were audio-recorded, each lasted 30-40 minutes.
All transcripts were read and analyzed, and relevant passages were highlighted. 
We applied an inductive approach to coding \cite{boyatzis1998transforming}, whereby themes are drawn from the raw data. Inductive coding was chosen to account for the breadth of experiences presented by our participants.

\subsection{Contextual Inquiry}
We conducted classroom observations with four male lecturers experienced in live coding from our university's computer science department, each covering two 45-minute lectures.
The lectures we observed covered topics introduction to programming in Python and C++, Data Science \& Machine Learning.
The study involved classroom observation and a semi-structured interview to understand their experience and mental workload during live coding sessions.
Teaching materials were collected beforehand, and instructors' behaviors and student interactions were noted. 
Data collection included field notes from two observers, audio/video recordings (when permitted), teaching materials (slides, code), and classroom artifacts (whiteboard snapshots). These were analyzed to identify recurring patterns and themes. 
Follow-up interviews were analyzed using the same coding method as the formative study.

%% file: sections/04_findings.tex
\section{Findings}

\subsection{Motivations (RQ1)}
\subsubsection{Improve Comprehension and Pacing.} 
Live coding helps instructors slow down the lecture speed, enabling students to follow the steps and understand the code, as explained by P1, \isay{There is a danger of when I show them a picture or slide, and I just continue, then I go too fast. If I do live coding, I cannot speed up it more than just writing the code
}.
This deliberate pace, combined with the ability to effectively clarify concepts through visual output, foster better comprehension.

\subsubsection{Adaptivity.} 
Live coding allows instructors to address student queries and adapt teaching to student needs in real-time. 
P5 emphasized the importance of being \isay{flexible enough to go into new paths} is the key to a good live coding session. 
Others highlighted similar benefits, like the ability to adjust solutions to address errors (P4) and the flexibility to modify the program directly to demonstrate concepts in response to student questions, unlike static slides (P8).

\subsubsection{Engagement.} Participants reported higher levels of student engagement and interaction during live coding sessions compared to traditional lectures, P4 noted, \isay{I found out that if I type in front of them, it makes them feel less pressure to raise their hands and interrupt my typing and ask the questions at that specific point}.
Instructors also mentioned that they got more specific and eager questions from students.
These observations align with prior research showing increased student interaction and engagement during live coding sessions \cite{iticse24-comparing}.

\subsubsection{Instilling Good Programming Practices.} 
Live coding was also seen as an opportunity to model and instill good programming habits.
P2 remarked, \isay{When I show students that I transform very unreadable code into readable code, I notice that in the next submissions, they follow these practices.} 
Similarly, P4 observed that live coding improved students' adherence to coding conventions and best practices, such as proper formatting and spacing.

\subsection{Implementations (RQ2)}
\label{sec:implementations}

\subsubsection{Preparation}
Contrary to what \cite{ace21-qualitative} reported that instructors often incrementally develop their program in a step by step manner, participants teaching advanced programming topics (e.g. recursion) consistently prepared full code solutions for live coding sessions prior to class to help reduce in-class stress (P1, P2 for university students, P3, P4, P6-P8), which can lead to preparation overhead (P2).
P7 and P8 mentioned strategies like leaving part of the important code cells in notebooks for live coding. 
In contrast, instructors teaching simpler topics (e.g. turtle programming) often adopted a spontaneous approach. 
P2 and P5 avoided preparing complete code in advance, as the simplicity of the tasks allowed them to code from scratch, better present their original thought processes. 
P5 described starting with only a skeletal framework \isay{I wouldn't have a fixed set of commands ready, I would put myself in the situation of the students that they would be starting from a blank slate}. 
Instructors described additional preparation tips, including using bullet points or printed solutions as references (P1), 
outlining key exercises or concepts to introduce (P5), and using ChatGPT to help generate exercises (P1).

\subsubsection{Procedure}
Our finding aligns with prior work \citet{ace21-qualitative} that documented the detailed live coding process in classrooms.
In short, in lectures (P6-P8), live coding typically lasts 5--10 minutes, either to demonstrate a concept or to motivate students, as instructors need to allocate time for additional topics. 
In contrast, live coding during exercise sessions tends to be more flexible, allowing for extended demonstrations and hands-on practice. 
These sessions often follow a structure where TAs first recap key concepts using slides, then allow students to work on problems independently before demonstrating solutions.

\subsubsection{Students' Activities}
Although none of the instructors required students to type code during exercises, many (P2 and P4) observed students actively coding alongside live demonstrations. 
In P6-P8's lectures, students were noticeably more engaged during live coding compared to slide-based presentations. 
They collaborated by pointing at each other’s screens to explain code or programming concepts. 
Students also asked more questions during live coding sessions than slide presentations, which aligns with previous findings \cite{iticse24-comparing}. 
The heightened interactivity motivated presenters to conduct additional testing in response to student queries. 
However, we observed a challenge in large classrooms: students found it difficult to call out the next line of code when instructors sought input. 
This highlights a limitation of verbal interaction in large classrooms, where the physical layout and class size can hinder effective communication.

\subsection{Obstacles (RQ3)}

\subsubsection{Decline of Engagement}
Live coding, while effective in demonstrating programming concepts, often struggles to maintain students' attention due to two common challenges.
First, the time-intensive nature of the activity can cause students to lose interest if the session becomes overly lengthy.
Second, instructors frequently become overly absorbed in coding tasks, such as debugging, which disrupts their connection with the audience.
As P2 noted, \isay{When the code get complex, I tend to focus too much on the debugging and I lose the audience}. 
These challenges echo findings from previous studies on live-streamed programming sessions \cite{vlhcc19-exploratory}.
To counteract these challenges, participants have shared techniques to engage students during live coding: asking for input and intentionally making mistakes. 
P6 noted that they only make deliberate mistakes when the code is simple, as more complex errors can confuse students and hard for themselves to manage. 
Additionally, P2 and P4 mentioned inviting students to the front of the class to present their code, which further fosters engagement by actively involving them in the process.

\subsubsection{Unpredictability}
Live coding is an improvisatory practice \cite{blackwell2022live} that often lacks a predefined structure \cite{ace20-live-static}. 
This flexibility allows instructors to go off-road, follow students’ interests, and explore additional concepts beyond a pre-prepared presentation \cite{iticse24-comparing}. 
However, such improvisation also introduces unpredictability, as live demonstrations may deviate from the plan, occasionally resulting in errors or unexpected outcomes.

Although most instructors prepare precise scripts (P1-P4, P6-P8) or code skeletons (P5) beforehand, deviations from these plans are common, often leading to unintentional mistakes. 
For example, P8 shared, \isay{I have the tendency to stray away from what I have prepared}.
These deviations are often a result of dual task of managing the class simultaneously while coding, and constantly referring to notes can slow down the flow. 
This reflects the tension between maintaining a smooth flow and referencing notes \cite{brown2018ten}.
Unpredictability is further amplified by student engagement, which introduces variations and disrupts time management. 
Instructors can adapt on the fly to accommodate student mistakes, which requires improvisation.
Such adjustments can deviate significantly from the original plan, and can further increase the likelihood of errors.
As P3 explained, differences in variable names and code structure can lead to unexpected deviations.
Debugging adds another layer of uncertainty, as instructors must first understand students’ reasoning before guiding them toward the correct solution, which often extends beyond the initial plan.
P1 reflected on this challenge, sharing that they rarely keep track of time during these interactions, \isay{I end up doing something, them asking questions, and it goes on, and I kept going until they understood everything}.

These factors--deviations from prepared plans, on-the-fly adaptations, reliance on memory, and student interactions--all contribute to the inherent unpredictability of live coding sessions. 
While this flexibility can enhance engagement and learning, it also demands considerable adaptability and real-time decision-making from instructors.

\subsubsection{Mental Stress}
Instructors reported high levels of stress caused by various reasons, including experience level, cognitive constrains of thinking aloud, multitasking process, and psychological stress when presenting publicly.

\textbf{Experience Level:} 
When instructors first began using live coding, they may have high stress levels. 
But over time, it decreased as they gained experience. 
P1, the least experienced instructor, described their initial anxiety: \isay{So when I did at the first time, I was very nervous that I couldn't do it, and I would make mistakes, and I wouldn't be able to find the mistakes.} 
This nervousness sometimes manifested physically, \isay{if I'm sometimes very nervous, I might like start shaking a little bit with my voice, and be blocked for a few seconds. Just don't see how to continue.} (P1)
More experienced instructors echoed similar sentiments but emphasized that familiarity with the format reduced the mental load over time. \isay{In the beginning, it was quite a lot of mental load to do live coding, I needed to think how to get it to work, what if I got it wrong and I don’t make typos. With more live coding sessions, the mental load is decreasing, but it’s still above the mental load of the normal lecture style.} (P7)
All lecturers in contextual inquiry rated live coding classes as more mentally demanding than normal classes except P8, the most experienced participant. 
P8 shared approaches such as frequently executing and verifying the result, or having a printed solution to help with unexpected mistakes.

\textbf{Think Aloud:} 
Thinking aloud while typing is a significant source of mental stress, as previously observed in technical coding interviews. 
\citet{esec20-stress} reported that being watched while thinking aloud lowers technical interview performance. 
P4 mentioned that this is manageable in simpler contexts, such as high school lessons.
In university level, this is unfeasible due to the increased complexity and length of the code. 
\isay{It was kind of possible to code and talk at the same time. It's not very challenging (in high school), but in university level, there's no chance of me doing that.} (P4) 
To manage this, P4 preferred to separate explanation and coding: \isay{I first explain to them and then I showed it to them.} 
Another challenge is the mismatch between spoken explanations and written code. 
P2 highlighted this difficulty: \isay{When I use the blackboard or slides, I write the same words as I say or the same words appear similarly on the slides. But when I say we now make a loop, I have to write something completely different than what I'm saying.}

\textbf{Multitasking:}
The multitasking nature could be another source of mental stress \cite{esec20-stress}. 
Novice instructors often feel overwhelmed by the amount of simultaneous activity in a classroom \cite{corcoran1981transition}. 
All of the TAs emphasized that live coding is more mentally demanding than conventional lectures, irrespective of the amount of code involved. 
The challenges include slow typing speed, managing multiple devices, and maintaining awareness of students' engagement.

\textbf{Fear of Making Mistakes:}
Spontaneous lectures are invigorating but carry the risk of mistakes that could confuse students and harm the instructor's confidence \cite{technique02}. 
Instructors often experience heightened anxiety during live coding sessions due to this fear, which adds significant pressure even for experienced teachers. 
P4 highlighted the importance of avoiding errors to prevent misguiding students. 
Similarly, P1 mentioned the challenges of explaining concepts under pressure, \isay{When you do live coding, you do it on the go. Mistakes can happen. You might forget something. You might explain something, not as you wanted it to, because you're nervous.} 
while P2 reflected on the stress of making mistakes, particularly as a new teacher. 
P3 noted the mental demands of live coding compared to regular lectures, and P6 emphasized the need for clarity when solving student problems, as hesitation or errors can confuse learners.

\textbf{Stage Fright:}
The public nature of live coding creates additional psychological pressure. 
P6 described the impact of being observed by a large audience: \isay{The feeling of being observed by hundreds of students creates a psychological pressure. There was laughter and students whispering, those things psychologically affected me.} 
This pressure also manifests as self-doubt, with P6 noting: \isay{There’s a little voice in my head that says, `Did I make a mistake, am I missing something, did they find a solution that I did not see', and that clatter in the head makes it hard to teach.}

\subsubsection{Time Pressure}
Live coding demands more time from instructors, both in preparation (as discussed in Section \ref{sec:implementations}) and in presentation. 
Compared to static code presentations, live coding sessions can take up to twice as long, as instructors must write and test code, explain their thought processes, and interact with students \cite{iticse24-comparing, ace20-live-static}. 
This sentiment was echoed by participants (P2-4, P7).

The improvisational and unpredictable nature of live coding further complicates time management. 
Instructors often struggle to accurately estimate durations. 
P2 reflected: \isay{I used to do a lot of time management, say I use 15 minutes for this, 20 minutes for this but I just saw that these estimates are so unrealistic because it depends on how many questions people ask.} 
Furthermore, bugs introduces additional delays, disrupting the flow of the lecture. 
P2 elaborated: \isay{I think the live coding always takes a lot more time than anything else because I have to write the code and students ask, then I make a mistake, something doesn't execute, then I have to check what it was.} 
P4 highlighted how spending too much time on debugging could reduce student engagement: \isay{This makes you feel you're wasting the time of your students and the students are getting bored in some way I feel that stresses me and that does make me not as likely to do live coding in university students because I know that it can easily be boring because it takes too long.} 
This worry aligns with previous research \cite{ace21-qualitative}, which noted that live coding’s slower pace and frequent interruptions can make it less time-efficient compared to static code examples.

%% file: sections/06_discussion.tex
\section{Discussion}
\subsection{Design Implications}

\subsubsection{Tailoring IDEs for Educational Live Coding} 
Tailored IDEs and real-time code sharing tools can significantly improve the effectiveness of live coding in educational settings.
Participants highlighted the challenges with current code editors such as PyCharm, IntelliJ, and Visual Studio Code. 
These tools, while powerful, often have overly complex interfaces that demand significant effort from students to navigate. 
Simplified IDEs with minimalist interfaces can reduce distractions and cognitive load while retaining essential functionality. 
Features like automatic syntax correction, guided code templates, and streamlined debugging workflows can further support instructors.
In addition, incorporating real-time code sharing features can improve engagement and interactivity in live coding sessions.
Existing tools such as Live Share\footnote{https://visualstudio.microsoft.com/services/live-share/}, Code With Me\footnote{https://www.jetbrains.com/code-with-me/}, and Replit\footnote{https://replit.com/}, allow multiple developers to browse and edit code simultaneously. 
However, they were not specifically designed for classroom use and often lack features like access control for student edits, which is critical in large classrooms with hundreds of students, or mechanisms for better code communication, such as allowing students to highlight code, suggest edits, or ask questions directly within the IDE.

\subsubsection{Adaptive Content Presentation.} The Improv IDE extension, which synchronizes code blocks and outputs with presentation slides, is one of the few efforts to help instructors manage their cognitive load during live coding presentation \cite{lats19-improv}. They lowered cognitive load by minimizing context switching and made it easier to fix errors on-the-fly. However, Improv is limited in its ability to adapt dynamically to the live classroom environment. It lacks features that allow instructors to respond to unanticipated student questions or adjust content based on real-time feedback. Future systems could incorporate AI-driven recommendations to help instructors manage unexpected scenarios during live coding. For instance, language models could suggest relevant code snippets or alternate examples based on student queries, helping instructors adapt their teaching on-the-fly. AI could also assist in debugging by identifying potential errors and offering step-by-step solutions in real time, reducing the cognitive burden on instructors. Moreover, integrating models' generative abilities into these tools could enable automatic transcription of verbal explanations into structured notes, allowing students to follow along more effectively.

\subsection{Limitations}
This study has several limitations to consider.
The small sample size, particularly in the contextual inquiry phase, may limit the generalizability of our findings. 
Future research should involve a larger and more diverse group of instructors across different institution. 
Additionally, our reliance on self-reported data from instructors may introduce biases. Future studies could incorporate more objective measures of cognitive load and stress, such as physiological data or performance metrics. Also, further research is needed to investigate the long-term effects of live coding on student learning outcomes and to develop strategies for mitigating the challenges identified in this study.

%% file: sections/07_conclusion.tex
\section{Conclusion}
Live coding is a powerful pedagogical technique that offers numerous benefits for programming education, including increased student engagement, deeper understanding of concepts, and the development of good programming practices. However, instructors face significant challenges when implementing live coding in the classroom, particularly related to its unpredictable nature, the mental stress it induces, and the limitations imposed by time constraints and available tools. Addressing these issues could improve its effectiveness by developing simpler, more intuitive IDEs tailored to the live coding context, creating real-time collaboration tools dedicated to live coding classrooms and incorporating language models to help with adaptive content presentation. Future research should explore the development of dedicated tools for live coding and examine their impact on teaching and learning experiences.

%% file: sections/08_ack.tex
\begin{acks}
We would like to express our gratitude to Dieter Schwarz Foundation for funding this research.
We also sincerely thank Carlos Cotrini Jimenez, Aaron Zeller, and Lukas Mast for their valuable contributions to the study, including data collection and insightful discussions.
Finally, we appreciate the time and effort of all participants who took part in our study. Their insights have been invaluable in shaping our findings.

\end{acks}